# When Screen Time Isn't Screen Time

## Tensions and Needs Between Tweens and Their Parents During Nature-Based Exploration

SABA KAWAS

University of Washington, Seattle WA 98195, skawas@uw.edu

NICOLE S. KUHN

University of Washington, Seattle WA 98195, nskuhn06@uw.edu

KYLE SORSTOKKE

University of Washington, Seattle WA 98195, kjs88@uw.edu

EMILY E. BASCOM

University of Washington, Seattle WA 98195, embascom@uw.edu

ALEXIS HINIKER

University of Washington, Seattle WA 98195, alexisr@uw.edu

KATIE DAVIS

University of Washington, Seattle WA 98195, kdavis@uw.edu

We investigated the experiences of 15 parents and their tween children (ages 8-12, n=23) during nature explorations using the NatureCollections app, a mobile application that connects children with nature. Drawing on parent interviews and in-app audio recordings from a 2-week deployment study, we found that tweens' experiences with the NatureCollections app were influenced by tensions surrounding how parents and tweens negotiate technology use more broadly. Despite these tensions, the app succeeded in engaging tweens in outdoor nature explorations, and parents valued the shared family experiences around nature. Parents desired the app to support family bonding and inform them about how their tween used the app. This work shows how applications intended to support enriching youth experiences are experienced in the context of screen time tensions between parents and tween during a transitional period of child development. We offer recommendations for designing digital experiences to support family needs and reduce screen time tensions.

CCS CONCEPTS • **Human-centred computing** → **Empirical studies in HCI;** *Empirical studies in ubiquitous and mobile computing and design*; • **Social and professional topics** → *Early Adolescents*.

Additional Keywords and Phrases: Child-computer interactions, early adolescence, tweens, parent-child relationship, families, smartphones, outdoor mobile technologies, nature-based explorations

**ACM Reference Format:**

# 1   Introduction

Tweens are characterized by a critical stage in development as they transition from childhood into adolescence [51,61]. They begin to separate from their parents, acquiring their sense of autonomous self, and they begin to spend more time with their peers [51,61]. Yet, tweens continue to rely on their parents for support and guidance through this transition [13,51,61]. In the midst of this shift, when tweens are negotiating their relationship with their parents and asserting their autonomy, many tweens receive their own smartphone [69]. As they start experiencing their autonomy and form their identity, they turn to their devices as portals for exploration and to connect with their peers [5,37]. Researchers reported that the phone has become a source of tension in parent-tween relationships [4,15,23]; some refer to the phone as a "transitional object" accompanying tweens' development from child to adolescent [15].

Popular media and common wisdom often portray screen time and technology use by tweens as a cause for concern [39]. Prior research describes the tension around technology use in families [1,4,15,43], parents' reactions to their children's technology use [23,60], and parent-teen perspectives on technology rules [23,34]. Parents of tweens report experiencing guilt and concern about the fast pace of their tweens' technology adoption and constant use [23,47,60]. Parents often feel the pressure of managing their tweens' and teens' screen time, including what they are doing with their devices [4,15]. This pressure is particularly salient during early adolescence, when many parents make the initial decision to give their tween a smartphone [15,21,59].

In addition to condemning screen time and smartphones as the cause of parent-tween relationship tensions, increasingly technologies are portrayed as harmful to our well-being [55,56]. In particular, screen time is often blamed for decreasing tweens' time spent outside in nature [40,42]. Research shows that time spent in nature plays a pivotal role in supporting children's learning, attention, physical health, and mental and emotional wellbeing [7,16,26,33,57]. Furthermore, nature-based activities and meaningful experiences in nature promote children's interest in natural environments, wildlife and plant species [9,10]. Yet, children today are spending less time in nature than previous generations, due to a variety of factors, such as fewer opportunities to access outdoor spaces in urban areas, parental safety concerns, busy schedules, lack of interest, and increased screen time [35,40,54,67,68,70]. A growing body of research demonstrates how mobile technologies can actually support children's positive, fun experiences outdoors, increase children's time outdoors, and support their engagement with and learning about nature [2,25,28,31,32,38,65,66].

Despite technology-related tensions in families, research also shows how technology can bring families together around shared interests through joint entertainment, play, and learning. The concept of joint media engagement (JME) illustrates the active, collaborative nature of children's technology use when it occurs in the context of family interactions and shared time [53]. Research on families' JME shows how it can support families' learning, playing, and co-creation of media content [3,36,46,49,52,53]. Recently, HCI researchers have studied JME in the context of outdoor spaces [49], specifically when designing for families and their children to support joint gaming and digital play outside. Location-based mobile games such as Pokémon GO and Ingress can motivate children and their parents to spend time outside and engage in enjoyable, fun activities together [29,49]. However, location-based mobile games were not designed specifically to encourage nature explorations by children and their families; moreover, spending time outdoors while searching for augmented reality (AR) creatures is quite different from focusing deliberately on nature interactions.

In prior work, we designed *NatureCollections*, a mobile application that connects tween children to nature and encourages them to go outside to collect and curate collections of nature photos [30]. The app allows children to classify and describe their nature photos and complete challenges to earn badges. In a prior evaluation study, the NatureCollections app successfully promoted children's curiosity of and connectedness to their natural surroundings [31]. The app succeeded at engaging children in enjoyable, meaningful nature explorations. Additionally, the app design is positioned to encourage nature-based interactions between tweens and their parents due to the app's emphasis on engaging youth in personally relevant activities, supporting focused attention on nature, encouraging social interactions (including with parents and siblings), and providing opportunities for continued joint engagement with nature [30]. In the current work, we are interested in understanding parents' experiences, perspectives, and their family needs around using NatureCollections to support joint family engagement with the app in the context of tweens' nature-based explorations and ways to increase family time spent outdoors. This work is situated in the broader context of parent-tween tech-related tensions in the adoption of mobile applications such as NatureCollections by tweens and their families [4,15,23,34,43,59,60]. We ask the following research questions: **RQ1**: How do parents experience their tweens' NatureCollections app engagement in outdoor exploration during their tweens' transitional stage of tech use? **RQ2:** How could design be used to support family needs around tweens' outdoor exploration during a transitional period of tech use?

This study was part of a larger app deployment experiment to evaluate multiple dimensions of tweens' app use. In this study, we examine parents' and tweens' shared experiences using NatureCollections. We provided 23 tweens (including eight sibling pairs), ages 8-12, from 15 unique families a phone with the NatureCollections (NC) app installed. Over the course of two weeks, we audio captured tweens and their families' interactions with NatureCollections, using in-app anchored audio recordings [22]. Following the two-week period, we conducted follow-up interviews with the parents. We found that the NC app successfully engaged tweens in nature-based explorations, and their parents valued their joint family experiences in nature. At the same time, tweens and their parents experienced these app interactions in the context of existing screen-time tensions.

This work contributes: (1) empirical evidence from parent interviews and tweens' in-situ NatureCollections app use that describe tweens' and parents' experiences and perspectives on shared family nature activities; (2) insights on how parents and their tweens negotiate technology use during a transitional period of tween child-development; and (3) design implications to support joint family nature-based explorations and recommendations for designers to reduce family tensions around screen time.

## 2 RELATED WORK

### 2.1 Technology-Related Tensions between Parents and Tweens

Digital devices are an integral part of tweens' modern lives. As tweens undergo a transitional developmental period from childhood into adolescence, many of them receive their own phone, at the average age of ten [69]. Tweens' developmental changes are often accompanied by parent-tween relationship tensions. As tweens begin to explore separation from their parents and establish their independence, they begin to spend less time with them and more time with their peers [13,51,61]. Yet, tweens continue to depend on their parents for instrumental support and guidance through this transition [13,61]. During this phase, tweens negotiate the transition from unilateral parental authority to a parent-child relationship that is marked by a greater degree of cooperation and compromise [13,61]. With this transitional parent-tween dynamic, tweens' phones often become the center of tension in the parent-tween relationship [4,15]. In fact, prior research referred to the phone as a "transitional object" accompanying tweens' developmental stage [15]. Tweens turn to their phones as portals for exploration and engagement with their peers as they form their identity and experience their autonomy. The phone's role also extends to keep tweens connected to their parents, as they spend increasing amounts of time away from them [4,15,34].

Many parents are worried about their children's screen time and concerned with the amount of time their tween children spend on their mobile devices, playing video games, watching YouTube, and talking to their friends over social media platforms [71]. These concerns are amplified by society and media messaging that stresses the risks of children's screen time (e.g., [56]). Consequently, parents are faced with pressures to mediate their tweens' technology use and enforce screen time limits. Research has found that children's views about their family's technology rules are somewhat conflicted. Children wanted their parents to guide their technology use and teach them how to be responsible with their devices, but at the same time, they wanted their parents to stop controlling their technology use and let them do what they want with their devices [23]. Prior research on family mediation strategies found that screen time rules focused on when, where, and how much teen children can use their devices. Parents controlled which games, apps, and social sites their tweens are allowed to access. Parental meditation theory describes three parental technology mediation roles: parents as *co-users* of media with their tween children; parents as *monitors*, actively monitoring their tween children's technology activities; and *restrictive* parents who restrict their children's online access and interactions [12,41]. Researchers have found that parental mediation of children's screen time is more nuanced in practice and is influenced by parent-child relationship dynamics, as well as parents' confidence in their technical knowledge to understand and manage their tween's technology use. Informed by the critical role of parent-tween tech-related tensions in prior research in family technology adoption, this paper investigates the nuanced interactions in parent-tween relationship dynamics using the NatureCollections app during nature-based explorations.

## 2.2   Family Joint Media Engagement

In addition to studying family tensions around screen time and phones, a growing body of research has explored how technologies and digital media use can bring family members together and documented the benefits of families' co-engagement with digital media. These practices are commonly known as Joint Media Engagement (JME) [53].  When parents and children engage in conversations and meaning-making together, children understand media content better and how it fits their family values. Prior work examined JME in different contexts such as co-viewing (e.g., watching TV shows or movies) [53], play (e.g., video games) [52], learning (e.g., eBook readings) and augmented reality games (e.g., Pokemon Go) [49].

Joint Media Engagement research on parents' and children's shared game experiences found that virtual and in-person collaboration during and around video gameplay promoted social interactions [52]. These interactions allowed parents and their children to transfer and share knowledge around common interests [52,53].  Prior work found that access to devices created shared family moments during family time, encouraged family meaningful conversations around shared interests, and supported family collaboration and creativity [63,64]. Yu et al. described that families used their mobile devices to look up information to help family decision-making for activities during vacations and achieve consensus among members. Vacation photos that the family took on their devices also helped form positive experiences around shared memories and brought family members together [63]. Sobel et al. found that families that played Pokémon Go, a location-based mobile game, valued that the gameplay increased shared family time outdoors[49]. Parents also felt that playing the game with their children facilitated spontaneous conversations that led to family bonding experiences.

Recently, HCI researchers have been interested in innovating new technologies to bring family members together, from supporting family collaboration and creativity to exploring new forms of interactive communication [20,58,62,64]. Yarosh et al. designed ShareTable, a system that allows parents and children who are separated to interact remotely and play together [62]. Other work investigated tangible storytelling systems to support communication between grandparents and their grandchildren [58]. Ferdous et al. designed a system, TableTalk, that transforms personal phones into a shared display to enrich family mealtime interactions and experiences [20].  Across all these studies, researchers found that the design of digital experiences plays a meaningful role in fostering joint family engagement, nurturing family connections and shared positive experiences with technology. The current work extends our understanding of families' Joint Media Engagement needs by considering tween-parent interactions in a new context: nature-focused exploration.

## 2.3 Mobile-Based Technologies for Nature Explorations

Our work designing NatureCollections is informed by prior HCI research that harnessed the affordances of mobile technologies to engage children in nature-based explorations. This work has primarily focused on supporting children's social play and science inquiry (e.g., [8,28,50,65]) by leveraging mobile, tangible, augmented reality, and sensor-based features. Interactive games that bridge physical and digital experiences with screenless devices, such as RaPIDO [50] and Scratch Node [24], have been shown to promote children's outdoor social game activities and embodied interactions. Pervasive popular games such as Geocaching [48] leverage location-based mobile features to support players in locating hidden treasures and collecting rewards in their physical world. Other augmented reality games like Pokemon Go [49] and Ingress [29] overlay co-located character avatars in the virtual game onto players' physical surroundings. Players can locate, capture, and battle virtual characters found by navigating spaces in the real world. A key goal across all these game design efforts is to support physical activity in outdoor gameplay and social interactions with other players.

Another body of mobile technology design research aims to support children's science inquiry and nature-based learning in outdoor contexts. For instance, Tree Investigators [66] and EcoMOBILE [28] leverage mobile and augmented reality capabilities to guide learners during field trips. Learners' physical surroundings are augmented with a virtual overlay of images and information to support their science inquiry. iBeacons [65], GeoTagger [19], and Tangible Flags [11] share learning activities with children based on their location to relevant nature elements. Across all these projects, researchers have aimed to promote children's scientific observations and increase their interactions with peers during outdoor field explorations.

Beyond the focus on children's physical play and learning outcomes, designing technologies that promote family joint nature-based explorations remains underexplored in HCI. At the same time, there is growing evidence in the learning sciences field that emphasizes the role of parents in supporting their children's sense-making and scientific observation in outdoor settings [17,44]. Here, we present insights from parent-tween experiences with the NatureCollections app during nature-based explorations and examine how specific affordances and design choices can support family joint nature-based engagement.

## 3 Method

In this paper, we examine tweens' and parents' shared experiences using *NatureCollections*, an app that encourages children to go outside and connect with the natural world [30]. We recruited 23 tweens (including eight sibling pairs), ages 8-12, from 15 unique families to use NatureCollections for a two-week period. This usage was part of a larger experimental deployment study that investigated multiple dimensions of the NatureCollections app [31]. Here, we examine only tweens' and parents' joint media engagement with NatureCollections. Participant demographics for this group are shown in Table 1.

**Table 1:** Demographic characteristics of families and their tweens.

| Demographic Variable |       | NC App* |
|----------------------|-------|---------|
| Gender               | Girl  | 12      |
|                      | Boy   | 11      |
| Age                  | Age 6 | 1       |
|  | Age 8 | 2 |
|  | Age 9 | 8 |
|  | Age 10 | 9 |
|  | Age 11 | 0 |
|  | Age 12 | 3 |
| Race | White or Caucasian | 13 |
|  | Black or African American | 1 |
|  | Asian / Pacific Islander | 1 |
|  | Mixed | 8 |
| Household Income (US$) | Less than $25,000 | 1 |
|  | $25,000 to $49,999 | 0 |
|  | $50,000 to $74,999 | 2 |
|  | $75,000 to $99,999 | 1 |
|  | $100,000 to $149,999 | 4 |
|  | $150,000 or more | 7 |
| Parent's Education | High school graduate | 1 |
|  | Trade /Vocational | 3 |
|  | Associate's degree | 2 |
|  | Bachelor's degree | 5 |
|  | Graduate degree | 4 |
| Birth Order of Primary Tween in Sibling Pair | Older than sibling | 5 |
|  | Younger than sibling | 3 |
|  | Single tween participating | 7 |

15 Families : *n = 23 total tweens, 15 primary participants and 8 siblings

## 3.1 Participants

We recruited families that had at least one tween child between the ages of 8 and 12 years. We attempted to recruit a diverse sample with respect to race, household income, and education level using a variety of recruitment strategies. We distributed flyers at local libraries, schools, and community centers throughout the metropolitan region where the study took place. We also posted the study announcement via a campus-wide news post and shared it in a local magazine. Authors used their personal social media accounts to share the study and we posted it on local parent Facebook groups. We had 164 qualified families interested in the study. We divided the tweens' families into two groups based on whether they reported that tween owned a handheld smart device (e.g. smartphone or iPod touch) or not. We then emailed equal numbers of families from each group based on the order they had signed up for the study to schedule the initial interview for the deployment study. We contacted a total of 151 families to schedule the initial interviews, and it took 3 weeks to complete the recruitment for the larger study. We gave families the option to meet us on campus, located in the center of the city, or at their neighborhood library to encourage the participation of lower socioeconomic families. Our final sample skewed towards upper- and middle-class families; however, it mirrors the race distribution of the urban city where the study took place. Families also reported living in neighborhoods that were distributed across the city metropolitan area.

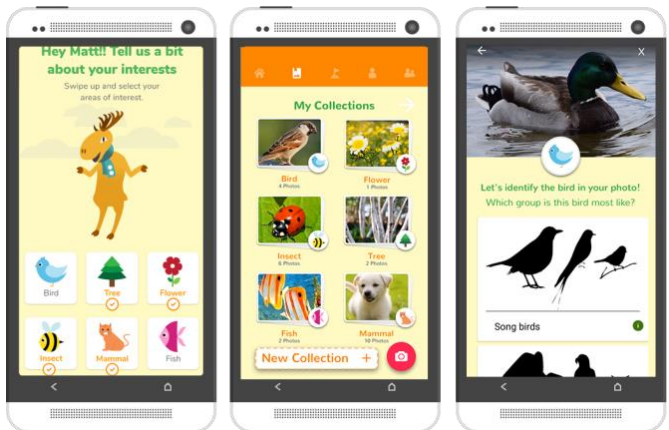

**Figure 1:** Screens of the Nature Go app 1: Onboarding "What are your interests?" 2: My Collections. 3: Classification

## 3.2   NatureCollections App

We designed the NatureCollections app in collaboration with children and guided by our interest-centered design framework [30]. NatureCollections encourages tween children to go outside to take photographs of nature, classify the plants, bugs and animals in their photos, and organize them into photo collections based on their species (e.g. insects and mammals). The app features are designed to spark tweens' interest in and connection to nature by supporting prolonged and direct interactions with nature. For example, the "Classification" feature, which provides a simple stepped visual prompt to guide the classification scheme for each photo Figure 1[3: Classification], allow tweens to direct their attention to the details of the natural elements in their surroundings. Similarly, the "Add Details" feature guides tweens with text-based prompts to enter descriptive information about their photos. These prompts lead tweens to closely examine the subject of their photo and reflect on the characteristics of the specific nature element. In prior work, we found that these features facilitated playful interactions around nature with parents, siblings, and peers [30]. Parents noticed an increase in their tween children's curiosity in their natural surroundings, and they observed that the app supported broader family engagement and conversations around nature elements [30].

To personalize tweens' NatureCollections experiences, the app records their current nature-related interests during the initial onboarding Figure 1[1: Onboarding]. Tweens are introduced to the app nature guide, a friendly moose character, who addresses children with their first name and prompts them to choose their interests. These interests are used to support child-driven interactions with nature by matching children's with the "Challenges" presented to them on the "HomePage." Tween children can organize their photo collections and create customized "My Collections" that tailor to their specific nature interests Figure 1[2: My Collections]. Children can view their photos taken to complete challenges and track their progress towards earning different badges. The app also includes a personalized "Profile Page" where tweens can track their accomplishments, including their challenges progress, photos collected, and earned badges. In addition, tweens can add their friends using a unique in-app username to "My Friends" and can view their friends' earned badges, number of photos taken, and completed challenges. We found that these personalized features extended tweens' engagement and increased social interactions with their siblings and peers, ranging from competing against to collaborating with each other to complete photo collections and challenges [31,32].

## 3.3   Study Procedures

The three-week experimental study consisted of an initial interview with both parents and their tweens, a one-week baseline period, and two-weeks of NatureCollections app use. Once families completed all the study procedures, parents and their tweens were invited to participate in final interviews to reflect on their family experiences using the app. During the first week of the study, families logged their tweens' daily activities and technology use, which was used to collect baseline data for each tween. This data is outside the scope of the current investigation and is unrelated to the study research questions posed here. Families then received a phone with the NatureCollections app installed to use for two-weeks. After this two-week period, we interviewed parents while their tweens engaged in an outdoor activity showing us how they had used the app during the study. Families that completed all parts of the study received US$75 gift cards, in which $50 was given after completing the final interviews. Our study was approved by the university Institutional Review Board and both parents and their tweens provided consent or assent to participate in the study.

In this paper we examined parent-tween experiences in the context of tweens' NatureCollections app use by analyzing: (1) parent self-reported descriptions of their experiences during exit interviews, and (2) tweens' and their families' real-time interactions when using the app, recorded using the Anchored Audio Sampling method (AAS)[22]. During the parent interviews, we used a semi-structured protocol and asked them about their family's and tweens' experiences with the NatureCollections app during their nature-based explorations. The length of these interviews ranged from 37 to 51 minutes ($M= 43$). All interviews were audio recoded and transcribed for data analysis. In-app AAS recordings were 3 minutes long and were captured during tweens' NatureCollections app use. AAS is a remote audio recording technique that is triggered seamlessly in response to a specific interaction to extract qualitative audio snippets during field deployment with children. These AAS recordings capture how children make sense of technologies and how they use them in their everyday life [22]. Each audio recording was randomly triggered either at the start, middle, or closing of a NatureCollections session to sample the full spectrum of a tween's app use experiences. The app displayed a notification on the app user interface that audio was being recorded. Recordings were stored locally on the device and then uploaded to our university servers once the device was connected to WiFi. At the end of the study, families had the option to delete some or all of the app recordings, though none of the families chose to delete any recordings. We captured a total of 704 recordings across all 23 tweens during the two-week period ($M= 29$ files per tween, $SD= 19$). Seven of the audio files were empty, indicating the participants did not say anything during the audio samplings. All audio files were transcribed verbatim.

## 3.4 Data Analysis

We took a joint inductive-deductive analysis approach to our qualitative data sets [14], the parent interviews and the app AAS recordings. Two researchers independently open coded 20% of the parent interviews following an inductive approach, and both researchers met regularly to discuss emerging themes and iterate on the codes for consistency [6]. Next, three researchers, including one who participated in coding the parent interviews, read through 20% of the transcribed AAS recordings from the phones, a total of 137 files. Following a similar inductive approach, the researchers coded the transcripts while writing memos on any new emerging themes, and discussed these themes in team meetings. Finally, following a deductive approach, all researchers looked through similarities across themes in both the parent interviews and the AAS recordings while iterating on the themes and adding any new codes to the codebook. We used Dedoose[*] to code both qualitative data sets, with one researcher coding the remaining parent interviews and the other researchers splitting the audio recording and coding them separately. The researchers met regularly to discuss the emerging themes and excerpts from the data.

## 4 RESULTS

Two overarching themes emerged from our data sources: (1) families' experiences of tweens' NC app use during nature-based explorations, and (2) concerns and tensions surrounding tweens' technology use during their transitional period of development.

---

[*] Dedoose (https://www.dedoose.com/) is a web-based research application for analyzing qualitative data in a collaborative asynchronous environment.

## 4.1 Families' Experiences of Tweens' App Use during Nature Exploration

### 4.1.1 How tweens engaged their parents and siblings around app use

Tweens engaged with their parents and siblings in joint nature-based explorations during their NC app use, often sharing their app activities and achievements with them. Tweens' app use increased their parents' attention to and engagement with nature elements in their surroundings. Additionally, tweens and their families worked together to make sense of nature elements. Some parents incorporated their pre-existing interests in nature activities and, at times, their knowledge around plants and other species while their tweens used the app.

Tweens shared their app photos, activities, and achievements with both parents and siblings during and after their app use. We heard one tween tell his mother:

*"Can I show you my favorite picture? I wanna show you my favorite picture. It's amazing, okay. ... I'm doing a night tree series"* to which she responded encouragingly, *"I like that one, it is a really good picture. Oh, that's a pretty one. Yeah, very good." AAS [T7 (boy, age 10)].\*\** T7's father commented, *"He was super jazzed about it. He was showing me every single picture he took that evening that he first tried it out. He had actually some really good pictures. It was cool." P7 [T7 (boy, age 10)].*

Tweens sought their parents' and siblings' help in taking photos for the app, with one tween who asked his father:

"*Right before you go to bed, do you think you could wake me up and take me outside, cause I wanna take a picture of the moon, cause one of my categories is moons?*" *AAS [T7 (boy, age 10)].*

Sibling pairs were each provided with phones for the study and engaged with one another, sharing photos and achievements while using the app collaboratively. A brother-sister pair talked about their photos and helped each other find nature elements:

*S5: "There's a lot of types of birds." And her brother asked "T5: Do you have a lot of birds [photos]?"*

*S5: "No. There is a lot of types of birds. Stop."*

*T5: "Let me see." Then the brother pointed to his sister:*

*T5: "There was a bird over there, look." AAS [T5 (boy, age 10), S5 (girl, age 8)].*

Many parents noticed that their tweens engaged in outdoor activities differently when using the app, with one mother, P8, sharing:

*"He never said anything to me, but he said, 'Oh, I need to take a picture of this marigold.' That was really hard to, but he had never, ever mentioned the marigold that we've had for six years, since before. Yes, it did actually help him interact and verbalize what he's seeing, definitely." P8 [T8 (boy, age 9), S8 (girl, age 12)].\*\**

Tweens' app use during regular family activities like going on walks, heading to the grocery store, and going out to dinner prompted an increase in their siblings' and parents' attention to nature and encouraged them to slow down, observe their natural surroundings, and engage in discussions about the photos they were taking. We heard a conversation between one tween and her mother:

*T12: "There's a snail right here."*

*P12: "Where? I don't want to step on him!"*

*T12: "He's right there. I'm taking a photo of it."*

*P12: "It's here? Let [your brother] get a picture of it." AAS [T12 (girl, age 12), S12 (boy, age 10)].*

This same mother shared with us about her two tweens*:*

*"I remember there was this bird on the fence that they were trying to get a picture of, and I remember we were just all sitting around for a while waiting for this bird to come back because it was flitting around. They were just super focused trying to get a picture of this bird. I can't remember if they actually got it [chuckles]." P12.*

Another parent, P14, mentioned:

---

\*\* T refers to tween, P refers to parent, S refers to sibling

> *"There was a day when we were coming home from her grandma's house and we were driving through the neighborhood. Then she yelled out, 'Oh, there's that bush! I remember that bush. I took a picture of that bush for the NatureCollections app and then I recorded it, why it was important to me.' I was like, 'Oh, I've never really noticed that bush.' [laughing] She was like, 'I know, now I'm always going to remember because I took a picture with it'." P14 [T14 (girl, age 9)].*

In some families, tweens' joint app engagement also supported family time outdoors. One mother noted: "*We definitely just took more random [family] walks, that the kids initiated just to take photos of nature.*" She further explained that her daughter and son do not often play together, but since they had the NC app: "*They were definitely interacting because of the app and with the app.*" She also observed her younger son develop an interest in something other than video games, like photography and nature (*P12 [T12 (girl, age 12), S12(boy, age 10)]*).

Tweens worked with their parents and siblings to make sense of nature elements, inspiring togetherness as they named plants in their photos, engaged in app activities, discussed nature elements in detail, and looked up nature-related app content. One mother, P2, shared:

> *"We were both arguing about what we can define, what the plant looks like, [laughs] … so we would both go back and forth all the time about what we would call this…because we don't really know what plant [it is]." P2 [T2 (girl, age 9)].*

We heard one tween, T15, and her mother working together to classify a tree using the app:

> *T15: "Mom, those trees back there are ferns, right?" …*
> *P15: "Fir. Douglas fir. There's some fir trees back there, and apple trees. A whole bunch of trees back there." And the mother continues:*
> *P15: "Needle. Fir trees have needles." …*
> *T15: "Fruit, fleshy berry-like things. So juniper is a tree?" …*
> *P15: "It is a type of tree. It's a bush tree." …*
> *T15: "I put it as a shrub." …*
> *P15: "Shrub, actually that's a better way to put it." AAS [T15 (girl, age 9)].*

Another parent mentioned how her tweens' app use:

> *"caused us to talk about things that maybe we wouldn't have talked about, leaves and the color of flowers and things like that" P3 [T3 (girl, age 9), S3 (girl, age 10)].*

While some tweens competed with their siblings for the highest number of photo collections, badges earned, and challenges completed, many tween sibling pairs also supported each other with other app activities like choosing photo names, adding photo details, and making sense of nature elements to classify photos. For instance, these two sisters helped each other describe a photo of a pine tree and figure out the tree type while going through the classification scheme in the app:

> *T4: "How would you describe the pine tree?"*
> *S4: "The pine tree is spiky but soft."*
> *T4: "Spiky but soft. Spiky but elegant."*
> *S4: "I got half of it. T4 I was alright typing "spiky but," but ok."*
> *T4: "Spiky but elegant. On a walk. I'm just going to type on a walk for where we found it.: It's a tree. Conifer. Is it needle or scale type?"*
> *S4: "Needle."*
> *T4: "Needle, yeah. Woody cones? Yeah, woody cones. Needles."*
> *S4: "No that one."*
> *T4: "Fir, spruce…Hemlock? Don't touch it."*
> *S4: "I won't. I'm not touching it." AAS [T4 (girl, age 10), S4 (girl, age 9)].*

Parents shared that the NC app brought families together during nature activities. For instance, one parent shared how she would regularly go on walks with her tween, but he would usually be scootering a couple blocks ahead of her. When he was using the NC app, in contrast, they walked together, looking at plants, talking about animals and even the crevices in the sidewalk. *"We were actually having a conversation, usually we don't"* P1 [ T1 (boy, age 10)].

Parents' nature-related interests such as gardening, walking, and hiking regularly provided context for tweens' NC app use. Some tweens also incorporated their parents' love of nature into their photos, with one parent, P13, describing:
> *"It was so pretty like just a sunset and how the trees were. It was just so pretty. He knows that my favorite part of the day is the sunset. He's like, 'I'm going to take this picture, mom.'"* P13 [T13 (boy, age 10)].

Parents also shared with their tweens their interest in specific nature elements, like teaching their tweens about different flower names and sharing details about their favorite flower as their tween took photos of it. One mother reported how she often visited their local P-patch with her tweens, noting how they engaged differently in these outings while they were using the NC app*:*
> *"It was the first time they had really looked around and asked specific questions about the plants around them considering, 'Hey, what are other people growing? What is this?'"* P12 [T12 (girl, age 12), S12 (boy, age 10)].

### 4.1.2    Parents' needs for engagement

After spending two weeks engaging and observing their tweens' NC app use, parents provided their insights on additional app features that would increase both their tweens' app engagement and their joint family nature-based explorations. Three main areas of need emerged: (1) app features to increase access to information about nature; (2) app activity sharing to support parent-tween engagement; and (3) new opportunities for family collaboration and competition. Many parents expressed a strong interest in having more app-supported opportunities for their tweens to increase their knowledge about nature. One parent, P7, suggested access to nature information via the app would be a key element for her to support her tween's use of the app:
> *"Having as much information about whatever you take a picture of, would be the most important thing that would get me to download [the app]."* P7 [T7 (boy, age 10), S7 (boy, age 6)].

Parents often shared that their tweens wanted the app to help them more as they were choosing names, classifying photos, or working on challenges. Many parents also wanted to use this additional information themselves to engage with their tweens in information seeking about nature and to support nature-focused family conversations. As one mother, P9, suggested: "*If it's a beetle, some parents might not know much about beetles but have casual like facts or research about it and you can have a conversation point.*" P9 [T9 (girl, age 9), S9 (girl, age 8)]

Parents shared a common desire to see how their tweens used the app and the photos they took, either through automatic activity reports or through tweens selecting photos to share. One mother, P8, shared her ideas for a reporting feature:
"*…like having some kind of a little report card. This is what your kid did today, or this week. This is how many pictures he took, very little synopsis of what kind of pictures they were; were they all plants? Were they all animals? … I think it'd be interesting to know what interests him, is it all flowers.. animals..mountain scenes..?*" P8 [T8 (boy, age 9), S8(girl, age 12)].

Parents largely expected this reporting to be facilitated electronically through the app, emails or text messages rather than by viewing the information on their tweens' phones. They also wanted the app to prompt their tweens to interact with them. as one parent, P3, suggested that tweens could:
> "*…get a notification that they've taken so many pictures and … see a notification like, 'Oh, you've taken 20 pictures. Go share them with the parent,' or, 'Let's talk with somebody about it.'"* P3 [T3 (girl, age 9), S3 (girl, age 10)].

Further, parents wanted to keep engaging their tweens by sharing comments on photos, titles, and classifications, and through a chat feature that would allow them to prompt their tween with questions. As one parent observed, this additional engagement through the app could serve as family conversation starters:
> "*…there's at least little bits I can pull out to have bigger conversations or uncover things I should know that otherwise if I was like, 'Tell me about your day,' I'd never get anywhere.*" P14 [T14 (girl, age 9)].

Numerous parents wanted even more app-facilitated, family-focused activities, with many expressing a desire for parents and extended family members to participate in collaborative and competitive activities with their tweens. Some also felt that having the app offer these challenges would more effectively engage their tweens in family activities:

> *"I know some kids that would have an issue with something being forced upon them from a parent, like, 'Go do this,' rather than the game actually prompting you to do it … a kid would be much more inclined to follow whatever incentive it is that the game actually promoted it, rather than the parent." P7 [T7 (boy, age 10), S17 (boy, age 6)].*

Parents shared many ideas for activities that would increase their tweens' family and app engagement, including daily goal setting and competitions for taking a specific number of photos. One parent, P1, shared how she envisioned competition with her tween:

> *"He's old enough now where he doesn't want me butting in on his challenges. He doesn't want me helping him. He wants to do everything himself, so he would never ask me, 'Go try to find a deciduous tree.' He would want to do it, and then show me that he found more than I did." P1 [T1 (boy, age 10)].*

## 4.2 Tweens' Transitional Tech Use

### 4.2.1 How the NC app fit into parents' screen-time rules

All parents indicated they have screen-time rules for their tweens, including rules about total screen time allowed per day, when and where tweens could use devices, parental screening of apps before downloading, and restrictions around chatting with friends on apps. Parents described the tensions they face around screen-time rules as their tweens transition between childhood and adolescence. A mother, P5, shared her experience with this tension:

> *"I'm kind of trying to just keep the 30-minute lockdown right now, for as long as I can, and that's all he knows. But I feel like that his being in fifth grade next year, the junior high kids start getting phones and there's less limitations on it, so I don't know, it's gonna be a whole different thing to kind of deal with." P5 [T5 (boy, age 10), S5 (girl, age 8)].*

Parents reported making exceptions to screen-time limits when technology use facilitated a valued family activity. They explained that they positively viewed technology engagements that support spending time outdoors, connecting with and learning about nature, social interactions with family, and creative activities over mindless tech use. One parent, P3, told us:

> *"It would be nice to be able to have something on a device, an activity that was more educational or help them see the world in a different way. Not just sit there. That's my biggest issue with the devices and the TV. You're just sitting there being a zombie. You're plugged in, and that's it." P3 [T3 (girl, age 9), S3 (girl, age 10)].*

Parents mentioned that they did not count their tweens' time on the NC app as screen time, and the app use therefore did not impact the amount of regular screen time their tween had on other devices. When we asked parents if this was due to participating in the study, they explicitly said it was not due to the study but rather because the NC app encouraged their tweens to go outside. One parent, P14, even suggested the NC app could be successfully marketed to parents with the tagline:

> *"Help balance your child's screen time, help your child spend more time outdoors with their technology."* This parent further explained: *"They are not even going to have to read much further, they'll just be like, 'I'll spend the US $3.99.'"*

Parents also identified other aspects of the NC app that they felt set it apart from the typical screen time they try to regulate:

(1) <u>Connecting with and learning about nature:</u> Parents expressed strong values associated with supporting their tweens' ability to connect with and learn about nature. One mother, P11, shared:

> *"I think all families, anybody that you ask, will tell you that the struggle is real, trying to push them away or pull them away from this video game, anxiety of winning competition, and having this relationship with people that you don't even see or know or have, there's no real connection. And I think given the alternative to have this connection with nature, I think it goes a long way." P11 [T11 (boy, age 12), S11 (boy, age 10)].*

(2) <u>Family interactions and activities:</u> Parents often remarked positively on the social interactions they observed when they had two children using the app together. A parent of a 12-year-old girl and a 10-year-old boy shared*:*

> *"Most of the screen time they use is a personal thing. They sit with YouTube or whatever, and it's just them and the screen. The NatureCollections app just encouraged more sharing, especially between the two of them. They would go out together, and they would just talk about what they were going to take pictures of. They would show each other their pictures, so that was definitely different." P12 [T12 (girl, age 12), T12 (boy, age 10)].*

(3) <u>Creative engagement</u>: Additionally, parents related other valued interactions to their tweens' NC app use such as creativity. A mother, P12, shared about her two children*: "They had a lot of fun taking pictures with the app. I thought that was good because-- Photography is a creative outlet, so I thought that was the positive use for technology to use it in that kind of way." P12 [T12 (girl, age 12), S12 (boy, age 10)].*

Parents also noticed that NC app features had prompted their tweens' imagination while interacting with natural elements. One parent, P2, shared an instance where she and her daughter spent some time in a garden and she noticed her tween:

> *"...had a whole different look of imaginations of how the fruits and the vegetables and the pictures. That was kind of fun to see, watch her imagination be a little bit- something that we normally wouldn't do. That she's using her imagination calling, labeling, and trying to describe flowers and things. That was kind of fun." P2 [T2 (girl, age 9)].*

Despite these valued outcomes, tensions still arose around phone use and the NC app. For instance, a few parents had family rules that restricted technology use outdoors. One mother shared:

> "*Because she's 9 and she doesn't have a phone, her go-to attitude about electronics is not to take them outside. As a family, she's not allowed to really take her iPad outside. It's like if you're going to go outside, part of why you're going outside is to disconnect from electronics, not take them with you*" P14 [T14 (girl, age 9)].

Another parent affirmed the general lack of familiarity among parents with the concept of tweens using technology outside. When she was asked about her child's use of technology outside during the study, she responded:

> *"I would always be confused by that question, because to me, spending time outside and technology don't go together, just because devices can get damaged. 'Cause when my kids play outside, it's all about being in motion." P11 [T11 (boy, age 12), S11 (boy, age 10)].*

When tweens took the phone to new outdoor spaces, they were often challenged with taking responsibility for it. One mother shared how she did not want to carry both of her tweens' phones for the study:

> *"I'm not carrying it, because I already have the dog's stuff, my own stuff, my own water .. they decided they wanted to take it and ... I remember my husband ended up carrying them." P5 [T5 (boy, age 10), S5 (girl, age 8)].*

### 4.2.2    Negotiating independence in phone and app use

Some parents shared that they did not spend much time helping their tween learn how to use the app after its initial setup. One parent commented:

> *"To be honest, I didn't really look at the app too much. I just let him do his thing. I didn't want to suggest too much to him, what he should do with this. I stayed away from that. He showed me pictures often of what he took, but that's about all I saw of the app." P7 [T7 (boy, age 10), S7 (boy, age 6)].*

Many parents mentioned that they and their tweens preferred this level of independent app use, with one parent stating her daughter's preference for independence using the app:

> *"She likes to do it mostly on her own. The only time we really had interactions if she had a question about something ...Then she was wanting a lot of interaction to try to help her figure out and that was towards the beginning. I think it was still fairly new to her, but after that she didn't ask for a lot of guidance with it. She just likes to do it on her own." P10 [T10 (girl, age 9)].*

At the same time, many parents also voiced concern about the safety measures that would be available to parents to monitor and manage tweens' social interactions. Numerous parents stated a desire to manage their tweens' friends on the app, with one father commenting:

> *"I think the ability to have a lot of control I think as a parent that's, I think, important. If she were to have her own device, we would want control over who she actually makes friend relationships with and it wouldn't be her decision alone. I would expect that to be on the other side as well." P6 [T6 (girl, age 9)].*

Parents also expressed concerns about their tweens' activities with friends on the app, including messaging and sharing photos:

> *"It's a double-edged sword. It would be nice to have, but then I can also see it getting out of hand, and maybe they start sharing pictures of all kinds of stupid random things...There would have to be some kind of parental control." P7 [T7 (boy, age 10)].*

The same parent suggested parental controls to limit the number of chats being sent to friends and regular reports to parents with the photos shared between friends. Additionally, numerous parents wanted to receive regular updates on their tweens' NC app use, including the amount of time they spent using the app. Several parents also voiced concerns over privacy:

> *"I would just not want any public online account sort of thing. If you could not do that, that would definitely be a plus" P12 [T12 (girl, age 12), S12 (boy, age 10)].* Some parents expressed a desire to limit their tweens' online presence: *"I am very concerned about their online presence. The trail, I don't want pieces of their address. Sometimes if they asked you where do they go to school? … I don't want my kids to put their real names." P9 [T9 (girl, age 9), S9 (girl, age 8)].*

Another dimension of negotiating independence in phone and app use related to the geographic limitations experienced by the tweens in this study, which impacted where they could use the NC app and their ability to venture out to new spaces without their parents. Parents reported that their tweens mostly used the app in their own yards and in their neighborhoods, close to their homes. When tweens used the app farther from home, they were largely accompanied by a parent, often going on walks, riding bikes, walking their dog, and running errands. One mother shared her tween's limited access to nature spaces: *"We don't have a neighborhood where he can go out on his own. It's our backyard or when I was close with him." P1 [T1 (boy, age 10)].*

Two siblings in the study discussed their geographic limits:

> T12: *"Okay, are we going? Okay, well bye!"* …
> P12: *"Don't go farther than, you know, don't go far"* …
> T12: *"What did she say our limit was?"* …
> S12: *"She didn't say what [our] limit was. I think our limit is up to the fence back there"* …
> P12: *"Geezz, I think the limit is any parking lot. Whoops! We just crossed one! L-O-L X-D."*
> *[T12 (girl, age 12), S12 (boy, age 10)].*

Sometimes, tweens tried to negotiate with their parents to go to new places, such as the zoo, a nearby beach, or on family hikes to explore and take photos of new nature elements.

## 5 DISCUSSION AND DESIGN IMPLICATIONS

In this study, we describe how parents experienced their tweens' NatureCollections app engagement in outdoor exploration and uncover family needs around app use. Our results showed that the NC app succeeded in engaging tweens in nature explorations and that parents valued family joint activities promoted by the NC app. We also found, however, that the digital experiences intended to connect tweens to nature and enrich their outdoor explorations were influenced by parents' screen-time rules and parent-tween negotiations around technology use. In our discussion, we draw from our empirical insights and prior work to present recommendations for designers to support family joint nature explorations. We also identify several opportunities to reduce parent-tween screen-time tensions during tweens' transitional period of development.

*Joint nature-based exploration:* Our results showed that the NC app facilitated Joint Media Engagement (JME) [53], by supporting family togetherness during nature-based activities and promoting behaviors that aligned with family values. This work extends our understanding of JME by considering these interactions in a new context: nature-focused exploration involving tweens who are experiencing a transitional stage of technology use. We uncovered families' needs that highlight parents' desire to support their tweens' autonomy while maintaining connection and shifting some of the burden of guiding their tweens to the app. Similar to prior JME research [20,63], we found that tweens' app activities sparked meaningful family interactions with nature. Parents, however, also expected the NC app to support deeper conversations with their tweens by providing access to nature-related knowledge and by sending them contextual information about their tweens' photos and app activities. Like Sobel et al., our results show that the NC app encouraged families to create more family bonding experiences, including spending more time together outdoors [49]. Parents and their tweens worked together while using the NC app to identify and learn about nature, consistent with prior insights about parents taking on new facilitation roles to support family social learning experiences [44]. Yet, parents in our study also wanted to take on this new role through their own parent version of the NC app, allowing them to remotely share photos, collaborate on activities, and complete family challenges. Additionally, parents wanted the app to facilitate a variety of joint family outdoor activities such as scavenger hunts and time- or location-based challenges.

*NC app use during a time of transitional tech use:* Tweens are in a transitional period of development, seeking to re-negotiate parental boundaries as they establish greater personal autonomy [13,61]. These dynamics can lead to tension between tweens and parents related to technology ownership and use [15,34]. On one hand, parents appreciated how the app encouraged their tweens' independence to initiate time outdoors without parental nudging, and they believed their tweens would be more willing to engage in these activities if they were suggested by the app rather than by them. On the other hand, tensions emerged around the integration of the NC app into parents' existing screen-time rules. Parents indicated that they normally enforced screen-time limits for their tweens' technology use. However, they made exceptions to these rules for the NC app, citing how the app supported valued family activities that engaged their tweens with positive outcomes like spending time outdoors, connecting with and learning about nature, family bonding activities, and creative activities over passive media consumption. Additionally, tweens negotiated geographic boundaries that determined where they could go without parental supervision, defining the limits of their outdoor app use. These struggles are similar to prior work examining parent-teen tensions related to the phone [4,15], but shaped in distinct ways by the specific context of outdoor nature exploration.

In addition to the needs identified for family joint nature-based engagement, parents expressed concerns about parental controls and safety concerns, including privacy, safety, and social connections with non-family members through the app. Parents desired a range of engagement options to address these concerns, mirroring parental mediation styles [12,41]. Some parents wanted to restrict who their tweens could connect with on the app and their level of social engagements; other parents wanted to monitor their tweens' overall app activity; and still other parents wanted to become co-users with their tweens with a parents' version of the app that integrated with their tweens' NC app. With these empirical insights in mind, we offer the following design recommendations:

## 5.1    Facilitate Digital Experiences that Mind the Context Gap

Our findings demonstrate that parents felt they were missing context around their tweens' NC app engagement, including app-related nature knowledge. Parents explained that lacking this context impacted how and to what extent they reacted and engaged with their tweens' nature-based experiences. Parents' engagement needs spanned the desire to learn what interested their tweens about the photos they were sharing with them, the type of nature element in focus (e.g., plant or bug), and the number of photos and badges their tweens collected. Parents expressed needing access to contextual nature information to increase their knowledge to be able to support their tweens in identifying the species in their photos and surroundings. Parents also desired nature prompts and questions to facilitate sense-making and deeper conversations about nature with their tweens.

Previous research in the learning sciences field has found that when parents support their children's observations and sense-making during family nature-based explorations, it facilitates children's development of scientific thinking and shapes their problem-solving skills [17,45]. These findings provide a design opportunity to facilitate digital experiences that bridge the context gap around app interactions and nature information to increase joint family nature-based engagement. Designers could consider supporting in-app audio and visual snap features to collect the context of tweens' app use. This data could be displayed alongside the photo metadata, such as location and photo details entered by the tween. The app could incorporate just-in-time features to recall photo-related nature information and nature prompts to support family conversations during their joint nature explorations.

## 5.2   Support Co-Located and Remote Family Activities

In our study, parents expected to jointly interact with their tweens' app activities both while being together and when being apart. Parents had competing demands and responsibilities, yet they desired the app to support continued engagement with their tweens' nature explorations, even when they were physically separated. Other parents desired features to support family activities during outdoor trips and even suggested having a parent version of the app, where parents and their tweens work together on family collections and challenges. Parents felt that these additional features would increase their joint family engagement around nature and facilitate meaningful family and nature interactions.

Prior research found that when parents and children engage in meaningful nature-based experiences together, it advances their children's connection to nature and supports their conservation values and environmental stewardship development [18]. Our work suggests that families can share nature-based experiences both when they are physically together and also when they are apart. These insights provide a design opportunity to combine co-located and remote digital family activities to support family togetherness in nature-based explorations. Designers could consider facilitating parent-tween remote connection and interactions around nature by supporting features for parents to engage digitally with their tweens' shared photos. For joint family activities, the app design could include co-play features that support collaborative and competitive nature-based challenges to increase family engagement and interactions with nature.

## 5.3   Create opportunities for technology to reduce screen time tensions

In addition to design needs to increase family joint nature-based engagement, we found that tweens' experiences with the NatureCollections app were influenced by how parents and tweens negotiate screen-time and technology-use tensions. Prior work identified that the sources of parent-tween tensions are influenced by the transitional period of development that accompanies tweens' technology use [13,51,61]. We found that tweens explored their autonomy and independence from their parents when using the NatureCollections app. Tweens negotiated geographic boundaries that limited their outdoor explorations of where they could go without parental supervision. We also observed in our study that parents' roles to mitigate their concerns about their tweens' privacy, safety, and social interactions with technology use matched the parental mediation roles of restrictive, active, and co-use [12,41].

Recent work investigating parental mediation roles has found that parental mediation roles are not discrete but fall on a spectrum that parents employ in different contexts [27]. These findings provide design opportunities to create technology that reduces screen-time tensions by supporting the diversity of parental mediation approaches and augmenting them with strategies that support tweens' autonomy needs. Facilitating tweens' sense of choice is central for tween autonomy development and proper separation from their parents on the path to adolescence and adulthood [13,61]. Designers could consider supporting parent-tween negotiations around screen time by facilitating features that: (1) guide clear limit setting; (2) elicit a meaningful rationale for limits; (3) mitigate conflict by acknowledging tweens' perspectives; and (4) provide choices and options. Designs could support features for tweens to self-direct boundary setting, in agreement with their parents, around which apps to use on their device and for how long by providing easy app drag-and-drop or feature selection options. In the context of nature-based exploration, the app design could support parents and tweens in jointly establishing geographic boundaries where tweens can explore. The app might send a notification to tweens when those boundaries are approaching. Additionally, designs could provide guiding prompts to parents to support conversations with tweens around their family's shared values and explanations for technology restrictions. These prompts could also engage tweens in reflective practices that encourage them to take ownership of their actions. In these ways, technology designs could promote meaningful input from both parents and their tweens and enable tweens' involvement in decision-making around their technology use.

# 6   LIMITATIONS AND FUTURE WORK

There are several limitations to our study data. Although we made efforts to recruit demographically diverse families, our participants were mostly from middle-to upper-class backgrounds. Although other racial and ethnicity demographics mirrored the urban city population where the study took place, our sample is not representative of the broader US population. Furthermore, our data is highly skewed toward mother participants representing the parents' perspectives. Although this study includes audio recordings from in-situ app use by the tweens, it does not include tweens' perspectives around their screen time and technology use.

Future research could work towards involving participants from culturally and socio-economically diverse families, as well as include greater representation of fathers. Additionally, future research could include tweens' perspectives on their lived experiences during this transitional stage of development, particularly around their transitional tech use and screen-time tensions that surround those experiences.

# 7   CONCLUTION

In this study, we explored parents' and their tweens' experiences during family joint nature engagement in the context of using the NatureCollections app for two weeks. NatureCollections succeeded at engaging tweens in nature-based explorations, and parents valued the activities that the app facilitated, including family shared nature experiences. At the same time, our results show that tweens' app experiences were influenced by screen-time tensions between parents and tweens. Parents desired a variety of functionalities to support them with their tweens' transitional technology use in addition to their family needs to increase engagement in nature. We present a set of recommendations for designing technology-based experiences to support family bonding in nature. We highlight opportunities for designers to promote tweens' autonomy, support positive transitional tween-tech use, and reduce family screen-time tensions.

## ACKNOWLEDGMENTS
We thank both tween children and their parents, for participating in our study. Special thanks to Mina Tari and Monica Posluszny for helping our team collect data. This material is based upon work supported by the University of Washington Innovation Award. This project was approved by the Institutional Review Board at the University of Washington (IRB ID: STUDY00002801).

## REFERENCES
1. Tawfiq Ammari, Priya Kumar, Cliff Lampe, and Sarita Schoenebeck. 2015. Managing Children's Online Identities: How Parents Decide What to Disclose About Their Children Online. In *Proceedings of the 33rd Annual ACM Conference on Human Factors in Computing Systems* (CHI '15), 1895–1904. https://doi.org/10.1145/2702123.2702325
2. Tetske Avontuur, Rian de Jong, Eveline Brink, Yves Florack, Iris Soute, and Panos Markopoulos. 2014. Play It Our Way: Customization of Game Rules in Children's Interactive Outdoor Games. In *Proceedings of the 2014 Conference on Interaction Design and Children* (IDC '14), 95–104. https://doi.org/10.1145/2593968.2593973
3. Rafael Ballagas, Thérèse E. Dugan, Glenda Revelle, Koichi Mori, Maria Sandberg, Janet Go, Emily Reardon, and Mirjana Spasojevic. 2013. Electric agents: fostering sibling joint media engagement through interactive television and augmented reality. In *Proceedings of the 2013 conference on Computer supported cooperative work* (CSCW '13), 225–236. https://doi.org/10.1145/2441776.2441803
4. Lindsay Blackwell, Emma Gardiner, and Sarita Schoenebeck. 2016. Managing Expectations: Technology Tensions Among Parents and Teens. In *Proceedings of the 19th ACM Conference on Computer-Supported Cooperative Work & Social Computing* (CSCW '16), 1390–1401. https://doi.org/10.1145/2818048.2819928
5. Danah Boyd. 2014. *It's Complicated: The Social Lives of Networked Teens*. Yale University Press, New Haven, CT, USA.
6. Virginia Braun and Victoria Clarke. 2006. Using thematic analysis in psychology. *Qualitative Research in Psychology* 3, 2: 77–101. https://doi.org/10.1191/1478088706qp063oa

7. Colin A. Capaldi, Holli-Anne Passmore, Elizabeth K. Nisbet, John M. Zelenski, and Raelyne L. Dopko. 2015. Flourishing in nature: A review of the benefits of connecting with nature and its application as a wellbeing intervention. *International Journal of Wellbeing* 5, 4. https://doi.org/10.5502/ijw.v5i4.449
8. Thomas Chatzidimitris, Damianos Gavalas, and Vlasios Kasapakis. 2015. PacMap: Transferring PacMan to the Physical Realm. In *Internet of Things. User-Centric IoT* (Lecture Notes of the Institute for Computer Sciences, Social Informatics and Telecommunications Engineering), 139–144. https://doi.org/10.1007/978-3-319-19656-5_20
9. Louise Chawla. 2015. Benefits of Nature Contact for Children. *Journal of Planning Literature* 30, 4: 433–452. https://doi.org/10.1177/0885412215595441
10. Louise Chawla and Victoria Derr. 2012. The Development of Conservation Behaviors in Childhood and Youth. *The Oxford Handbook of Environmental and Conservation Psychology*. https://doi.org/10.1093/oxfordhb/9780199733026.013.0028
11. Gene Chipman, Allison Druin, Dianne Beer, Jerry Alan Fails, Mona Leigh Guha, and Sante Simms. 2006. A Case Study of Tangible Flags: A Collaborative Technology to Enhance Field Trips. In *Proceedings of the 2006 Conference on Interaction Design and Children* (IDC '06), 1–8. https://doi.org/10.1145/1139073.1139081
12. Lynn Schofield Clark. 2011. Parental Mediation Theory for the Digital Age. *Communication Theory* 21, 4: 323–343. https://doi.org/10.1111/j.1468-2885.2011.01391.x
13. W. Andrew Collins and Laurence Steinberg. 2006. Adolescent Development in Interpersonal Context. In *Handbook of child psychology: Social, emotional, and personality development, Vol. 3, 6th ed*. John Wiley & Sons, Inc., Hoboken, NJ, US, 1003–1067.
14. Juliet M. Corbin. 2015. *Basics of qualitative research: techniques and procedures for developing grounded theory*. SAGE, Thousand Oaks, California.
15. Katie Davis, Anja Dinhopl, and Alexis Hiniker. 2019. "Everything's the Phone": Understanding the Phone's Supercharged Role in Parent-Teen Relationships. In *Proceedings of the 2019 CHI Conference on Human Factors in Computing Systems* (CHI '19), 227:1-227:14. https://doi.org/10.1145/3290605.3300457
16. Raelyne L. Dopko, Colin A. Capaldi, and John M. Zelenski. 2019. The psychological and social benefits of a nature experience for children: A preliminary investigation. *Journal of Environmental Psychology* 63: 134–138. https://doi.org/10.1016/j.jenvp.2019.05.002
17. Catherine Eberbach and Kevin Crowley. 2017. From Seeing to Observing: How Parents and Children Learn to See Science in a Botanical Garden. *Journal of the Learning Sciences* 0, 0: 1–35. https://doi.org/10.1080/10508406.2017.1308867
18. Julie Ernst. 2018. Zoos' and Aquariums' Impact and Influence on Connecting Families to Nature: An Evaluation of the Nature Play Begins at Your Zoo & Aquarium Program. *Visitor Studies* 21, 2: 232–259. https://doi.org/10.1080/10645578.2018.1554094
19. Jerry Alan Fails, Katherine G. Herbert, Emily Hill, Christopher Loeschorn, Spencer Kordecki, David Dymko, Andrew DeStefano, and Zill Christian. 2014. GeoTagger: A Collaborative and Participatory Environmental Inquiry System. In *Proceedings of the Companion Publication of the 17th ACM Conference on Computer Supported Cooperative Work & Social Computing* (CSCW Companion '14), 157–160. https://doi.org/10.1145/2556420.2556481
20. Hasan Shahid Ferdous, Bernd Ploderer, Hilary Davis, Frank Vetere, Kenton O'Hara, Geremy Farr-Wharton, and Rob Comber. 2016. TableTalk: integrating personal devices and content for commensal experiences at the family dinner table. In *Proceedings of the 2016 ACM International Joint Conference on Pervasive and Ubiquitous Computing* (UbiComp '16), 132–143. https://doi.org/10.1145/2971648.2971715
21. Arup Kumar Ghosh, Karla Badillo-Urquiola, Mary Beth Rosson, Heng Xu, John M. Carroll, and Pamela J. Wisniewski. 2018. A Matter of Control or Safety?: Examining Parental Use of Technical Monitoring Apps on Teens' Mobile Devices. In *Proceedings of the 2018 CHI Conference on Human Factors in Computing Systems* (CHI '18), 194:1-194:14. https://doi.org/10.1145/3173574.3173768
22. Alexis Hiniker, Jon E. Froehlich, Mingrui Zhang, and Erin Beneteau. 2019. Anchored Audio Sampling: A Seamless Method for Exploring Children's Thoughts During Deployment Studies. In *Proceedings of the 2019 CHI Conference on Human Factors in Computing Systems* (CHI '19), 1–13. https://doi.org/10.1145/3290605.3300238
23. Alexis Hiniker, Sarita Y. Schoenebeck, and Julie A. Kientz. 2016. Not at the Dinner Table: Parents' and Children's Perspectives on Family Technology Rules. In *Proceedings of the 19th ACM Conference on Computer-Supported Cooperative Work & Social Computing* (CSCW '16), 1376–1389. https://doi.org/10.1145/2818048.2819940
24. Tom Hitron, Itamar Apelblat, Iddo Wald, Eitan Moriano, Andrey Grishko, Idan David, Avihay Bar, and Oren Zuckerman. 2017. Scratch Nodes: Coding Outdoor Play Experiences to enhance Social-Physical Interaction. In *Proceedings of the 2017 Conference on Interaction Design and Children* (IDC '17), 601–607. https://doi.org/10.1145/3078072.3084331

25. Tom Hitron, Idan David, Netta Ofer, Andrey Grishko, Iddo Yehoshua Wald, Hadas Erel, and Oren Zuckerman. 2018. Digital Outdoor Play: Benefits and Risks from an Interaction Design Perspective. In *Proceedings of the 2018 CHI Conference on Human Factors in Computing Systems* (CHI '18), 284:1-284:13. https://doi.org/10.1145/3173574.3173858
26. Andrew J. Howell, Raelyne L. Dopko, Holli-Anne Passmore, and Karen Buro. 2011. Nature connectedness: Associations with well-being and mindfulness. *Personality and Individual Differences* 51, 2: 166–171. https://doi.org/10.1016/j.paid.2011.03.037
27. Hee Jhee Jiow, Sun Sun Lim, and Julian Lin. 2017. Level Up! Refreshing Parental Mediation Theory for Our Digital Media Landscape: Parental Mediation of Video Gaming. *Communication Theory* 27, 3: 309–328. https://doi.org/10.1111/comt.12109
28. Amy M. Kamarainen, Shari Metcalf, Tina Grotzer, Allison Browne, Diana Mazzuca, M. Shane Tutwiler, and Chris Dede. 2013. EcoMOBILE: Integrating augmented reality and probeware with environmental education field trips. *Computers & Education* 68: 545–556. https://doi.org/10.1016/j.compedu.2013.02.018
29. Pavel Karpashevich, Eva Hornecker, Nana Kesewaa Dankwa, Mohamed Hanafy, and Julian Fietkau. 2016. Blurring Boundaries Between Everyday Life and Pervasive Gaming: An Interview Study of Ingress. In *Proceedings of the 15th International Conference on Mobile and Ubiquitous Multimedia* (MUM '16), 217–228. https://doi.org/10.1145/3012709.3012716
30. Saba Kawas, Sarah Chase, Jason Yip, Joshua Lawler, and Davis Katie. 2019. Sparking Interest: A Design Framework for Mobile Technologies to Promote Children's Interest in Nature. . *International Journal of Child-Computer Interaction*.
31. Saba Kawas, Nicole S. Kuhn, Mina Tari, Alexis Hiniker, and Katie Davis. 2020. "Otter this world": can a mobile application promote children's connectedness to nature? In *Proceedings of the Interaction Design and Children Conference* (IDC '20), 444–457. https://doi.org/10.1145/3392063.3394434
32. Saba Kawas, Jordan Sherry-Wagner, Nicole Kuhn, Sarah Chase, Brittany Bentley, Joshua Lawler, and Katie Davis. 2020. NatureCollections: Can a Mobile Application Trigger Children's Interest in Nature? 579–592. Retrieved September 16, 2020 from https://www.scitepress.org/Link.aspx?doi=10.5220/0009421105790592
33. Lucy E. Keniger, Kevin J. Gaston, Katherine N. Irvine, and Richard A. Fuller. 2013. What are the Benefits of Interacting with Nature? *International Journal of Environmental Research and Public Health* 10, 3: 913–935. https://doi.org/10.3390/ijerph10030913
34. Ada S. Kim and Katie Davis. 2017. Tweens' perspectives on their parents' media-related attitudes and rules: an exploratory study in the US. *Journal of Children and Media* 11, 3: 358–366. https://doi.org/10.1080/17482798.2017.1308399
35. Rachel Tolbert Kimbro, Jeanne Brooks-Gunn, and Sara McLanahan. 2011. Young children in urban areas: Links among neighborhood characteristics, weight status, outdoor play, and television watching. *Social Science & Medicine* 72, 5: 668–676. https://doi.org/10.1016/j.socscimed.2010.12.015
36. Marina Krcmar and Drew P. Cingel. 2014. Parent–Child Joint Reading in Traditional and Electronic Formats. *Media Psychology* 17, 3: 262–281. https://doi.org/10.1080/15213269.2013.840243
37. Kresimir Krolo. 2010. Mizuko Ito et al., Hanging Out, Messing Around, and Geeking Out: Kids Living and Learning with New Media. *Revija za Sociologiju* 40, 2: 231–234.
38. Susanne Lagerström, Iris Soute, Yves Florack, and Panos Markopoulos. 2014. Metadesigning interactive outdoor games for children: a case study. In *Proceedings of the 2014 conference on Interaction design and children* (IDC '14), 325–328. https://doi.org/10.1145/2593968.2610483
39. Simone Lanette, Phoebe K. Chua, Gillian Hayes, and Melissa Mazmanian. 2018. How Much is "Too Much"?: The Role of a Smartphone Addiction Narrative in Individuals' Experience of Use. *Proc. ACM Hum.-Comput. Interact.* 2, CSCW: 101:1-101:22. https://doi.org/10.1145/3274370
40. Lincoln R. Larson, Rachel Szczytko, Edmond P. Bowers, Lauren E. Stephens, Kathryn T. Stevenson, and Myron F. Floyd. 2018. Outdoor Time, Screen Time, and Connection to Nature: Troubling Trends Among Rural Youth?: *Environment and Behavior*. https://doi.org/10.1177/0013916518806686
41. Sonia Livingstone, Kjartan Ólafsson, Ellen J. Helsper, Francisco Lupiáñez-Villanueva, Giuseppe A. Veltri, and Frans Folkvord. 2017. Maximizing opportunities and minimizing risks for children online: the role of digital skills in emerging strategies of parental mediation. *Journal of Communication* 67: 82–105.
42. Richard Louv. 2008. *Last Child in the Woods: Saving our Children from Nature Deficit Disorder.* Algonquin Books. Retrieved September 10, 2017 from http://edrev.asu.edu/edrev/index.php/ER/article/view/1196
43. Melissa Mazmanian and Simone Lanette. 2017. "Okay, One More Episode": An Ethnography of Parenting in the Digital Age. In *Proceedings of the 2017 ACM Conference on Computer Supported Cooperative Work and Social Computing* (CSCW '17), 2273–2286. https://doi.org/10.1145/2998181.2998218

44. Lucy R. McClain. 2018. Parent Roles and Facilitation Strategies as Influenced by a Mobile-Based Technology During a Family Nature Hike. *Visitor Studies* 21, 2: 260–286. https://doi.org/10.1080/10645578.2018.1548844
45. Lucy R. McCLAIN and Heather Toomey Zimmerman. 2014. Prior Experiences Shaping Family Science Conversations at a Nature Center. *Science Education* 98, 6: 1009–1032. https://doi.org/10.1002/sce.21134
46. Lucy Richardson Mcclain. 2016. Family Learning with Mobile Devices in the Outdoors: Designing an e-trailguide to Facilitate Families' Joint Engagement with the Natural World. Retrieved June 25, 2020 from https://etda.libraries.psu.edu/catalog/28747
47. Gustavo S. Mesch. 2009. Parental Mediation, Online Activities, and Cyberbullying. *CyberPsychology & Behavior* 12, 4: 387–393. https://doi.org/10.1089/cpb.2009.0068
48. Kenton O'Hara. 2008. Understanding geocaching practices and motivations. In *Proceedings of the SIGCHI Conference on Human Factors in Computing Systems* (CHI '08), 1177–1186. https://doi.org/10.1145/1357054.1357239
49. Kiley Sobel, Arpita Bhattacharya, Alexis Hiniker, Jin Ha Lee, Julie A. Kientz, and Jason C. Yip. 2017. It wasn't really about the Pokémon: Parents' Perspectives on a Location-Based Mobile Game. In *Proceedings of the 2017 CHI Conference on Human Factors in Computing Systems - CHI '17*, 1483–1496. https://doi.org/10.1145/3025453.3025761
50. Iris Soute, Tudor Vacaretu, Jan De Wit, and Panos Markopoulos. 2017. Design and Evaluation of RaPIDO, A Platform for Rapid Prototyping of Interactive Outdoor Games. *ACM Trans. Comput.-Hum. Interact.* 24, 4: 28:1-28:30. https://doi.org/10.1145/3105704
51. Laurence Steinberg and Susan B. Silverberg. 1986. The Vicissitudes of Autonomy in Early Adolescence. *Child Development* 57, 4: 841–851. https://doi.org/10.2307/1130361
52. Reed Stevens, Tom Satwicz, and Laurie McCarthy. 2008. In-game, in-room, in-world: Reconnecting video game play to the rest of kids' lives. *The ecology of games: Connecting youth, games, and learning* 9: 41–66.
53. Lori Takeuchi and Reed Stevens. The New Coviewing: Designing for Learning through Joint Media Engagement. 75.
54. Pooja S. Tandon, Chuan Zhou, and Dimitri A. Christakis. 2012. Frequency of Parent-Supervised Outdoor Play of US Preschool-Aged Children. *Archives of Pediatrics & Adolescent Medicine* 166, 8: 707–712. https://doi.org/10.1001/archpediatrics.2011.1835
55. Sherry Turkle. 2011. *Alone together: why we expect more from technology and less from each other*. Basic Books, New York. Retrieved December 14, 2018 from http://public.eblib.com/choice/publicfullrecord.aspx?p=684281
56. Jean M. Twenge. 2017. *IGen: why today's super-connected kids are growing up less rebellious, more tolerant, less happy-- and completely unprepared for adulthood (and what this means for the rest of us)*. Atria Books, New York, NY.
57. Vidar Ulset, Frank Vitaro, Mara Brendgen, Mona Bekkhus, and Anne I. H. Borge. 2017. Time spent outdoors during preschool: Links with children's cognitive and behavioral development. *Journal of Environmental Psychology* 52: 69–80. https://doi.org/10.1016/j.jenvp.2017.05.007
58. Torben Wallbaum, Andrii Matviienko, Swamy Ananthanarayan, Thomas Olsson, Wilko Heuten, and Susanne C.J. Boll. 2018. Supporting Communication between Grandparents and Grandchildren through Tangible Storytelling Systems. In *Proceedings of the 2018 CHI Conference on Human Factors in Computing Systems* (CHI '18), 1–12. https://doi.org/10.1145/3173574.3174124
59. Pamela Wisniewski, Arup Kumar Ghosh, Heng Xu, Mary Beth Rosson, and John M. Carroll. 2017. Parental Control vs. Teen Self-Regulation: Is There a Middle Ground for Mobile Online Safety? In *Proceedings of the 2017 ACM Conference on Computer Supported Cooperative Work and Social Computing* (CSCW '17), 51–69. https://doi.org/10.1145/2998181.2998352
60. Pamela Wisniewski, Haiyan Jia, Heng Xu, Mary Beth Rosson, and John M. Carroll. 2015. "Preventative" vs. "Reactive": How Parental Mediation Influences Teens' Social Media Privacy Behaviors. In *Proceedings of the 18th ACM Conference on Computer Supported Cooperative Work & Social Computing* (CSCW '15), 302–316. https://doi.org/10.1145/2675133.2675293
61. R. M. Wrate. 1986. Adolescent Relations with Mothers, Fathers and Friends. By James Youniss and Jacqueline Smollar. London: The University of Chicago Press. 1985. Pp. 201. £21.25. *British Journal of Psychiatry* 149, 6: 805–805. https://doi.org/10.1192/S0007125000122354
62. Svetlana Yarosh, Anthony Tang, Sanika Mokashi, and Gregory D. Abowd. 2013. "almost touching": parent-child remote communication using the sharetable system. In *Proceedings of the 2013 conference on Computer supported cooperative work* (CSCW '13), 181–192. https://doi.org/10.1145/2441776.2441798
63. Xi Yu, Gerardo Joel Anaya, Li Miao, Xinran Lehto, and IpKin Anthony Wong. 2017. The Impact of Smartphones on the Family Vacation Experience: *Journal of Travel Research*. https://doi.org/10.1177/0047287517706263
64. Nicola Yuill, Yvonne Rogers, and Jochen Rick. 2013. Pass the iPad: collaborative creating and sharing in family groups. In *Proceedings of the SIGCHI Conference on Human Factors in Computing Systems* (CHI '13), 941–950. https://doi.org/10.1145/2470654.2466120

65. Heather Toomey Zimmerman, Susan M. Land, Chrystal Maggiore, Robert W. Ashley, and Chris Millet. 2016. Designing Outdoor Learning Spaces With iBeacons: Combining Place-Based Learning With the Internet of Learning Things. Retrieved September 4, 2017 from https://repository.isls.org/handle/1/349
66. Heather Toomey Zimmerman, Susan M. Land, Michael R. Mohney, Gi Woong Choi, Chrystal Maggiore, Soo Hyeon Kim, Yong Ju Jung, and Jaclyn Dudek. 2015. Using Augmented Reality to Support Observations About Trees During Summer Camp. In *Proceedings of the 14th International Conference on Interaction Design and Children* (IDC '15), 395–398. https://doi.org/10.1145/2771839.2771925
67. 2011. Kids These Days: Why Is America's Youth Staying Indoors? *Children & Nature Network*. Retrieved January 13, 2020 from https://www.childrenandnature.org/2011/09/12/kids_these_days_why_is_americas_youth_staying_indoors/
68. 2018. The Great Indoors: Today's Screen-Hungry Kids Have Little Interest In Being Outside. *Study Finds*. Retrieved September 16, 2020 from https://www.studyfinds.org/great-indoors-screen-hungry-kids-video-games-going-outside/
69. Kids & Tech: The Evolution of Today's Digital Natives | Influence Central. Retrieved October 22, 2019 from http://influence-central.com/kids-tech-the-evolution-of-todays-digital-natives/
70. *The Nature Conservancy, New research reveals the nature of America's youth.* Retrieved September 11, 2017 from https://www.nature.org/newsfeatures/kids-in-nature/kids-in-nature-poll.xml
71. The Common Sense Census: Plugged-In Parents of Tweens and Teens, 2016 | Common Sense Media. Retrieved September 17, 2020 from https://www.commonsensemedia.org/research/the-common-sense-census-plugged-in-parents-of-tweens-and-teens-2016